\documentclass[letter]{aa}  

\usepackage{graphicx}
\usepackage{txfonts}
\usepackage{afterpage}
\usepackage{multirow}
\usepackage{txfonts}
\usepackage{graphicx}
\usepackage[outdir=./]{epstopdf}
\usepackage{natbib}
\usepackage{xcolor, color}
\usepackage[skip=2pt]{caption}
\usepackage{booktabs}
\usepackage{threeparttable}
\usepackage{hyperref}
\hypersetup{breaklinks=true, linkcolor=red, citecolor=blue, filecolor=cyan, urlcolor=magenta}
\usepackage{comment}

\begin{document}

   \title{
    The environmental imprint on molecular layering in the dusty streamer of M512
    }
   \subtitle{}

   \author{M. De Simone
          \inst{1,4}
          \and
          L. Cacciapuoti\inst{2}
          \and
          D. Capela\inst{3}
          \and
          E. Macias\inst{1}
          \and
          L. Podio\inst{4}
          \and
          A. Miotello\inst{1}
          \and
          A. Gupta\inst{5}
          \and
          J. Bae \inst{6}
          \and
          S. Grant\inst{7}
          \and 
          C. Espaillat\inst{8}
          }

   \institute{ESO, Karl Schwarzchild Str. 2, D-85478 Garching bei München, Germany
              \footnote{corresponding author: marta.desimone@eso.org}
         \and
             ESO, Alonso de Córdova 3107, Vitacura, Casilla 19001, Santiago de Chile, Chile
        \and
        SUPA, School of Physics \& Astronomy, University of St Andrews, North Haugh, St Andrews KY16 9SS, UK
        \and
             INAF, Osservatorio Astrofisico di Arcetri, Largo E. Fermi 5, I-50125 Firenze, Italy
        \and
             Department of Astronomy, University of Virginia, Charlottesville, VA 22904, USA
        \and
            Department of Astronomy, University of Florida, Gainesville, FL 32611, USA
        \and
            Earth and Planets Laboratory, Carnegie Institution for Science, 5241 Broad Branch Road, NW, Washington, DC 20015, USA
        \and 
        Department of Astronomy, Boston University, 725 Commonwealth Avenue, Boston, MA 02215, USA
            }
   \date{}

\abstract
{Protostellar streamers are elongated structures that channel material from larger scale onto disks, influencing their physical and chemical evolution. The M512 protostar in Orion/Lynds 1641 hosts one of the most massive and extended streamer discovered so far, offering a unique opportunity to study these processes. We investigate the morphology, chemistry, and origin of this streamer, and its potential impact on the protostellar disk. Using archival ALMA observations of C$^{18}$O, DCO$^+$, N$_2$D$^+$, and HCO$^+$, we compare their spatial distributions through moment maps and spatial profiles. The streamer shows clear chemical stratification: C$^{18}$O lies on the western side of the protostar, N$_2$D$^+$ is farther out to the east, and DCO$^+$ is in the middle. This suggests that the structure has been shaped by environmental effects rather than tracing a single coherent infalling flow, with only the densest gas near the protostar likely to accrete onto the disk. Overall, the bulk of the streamer reflects the physical and chemical imprint of the surrounding cloud, highlighting the importance of environmental shaping in interpreting streamer–disk connections and their role in disk growth.}


   \maketitle
%

\section{Introduction}

Recent interferometric observations have reached unprecedented spatial and spectral resolution, and sensitivity to detect rotational molecular line emission revealing complex and extended gas structure around young stellar objects. On scales from $\sim$150 au to 10000 au, elongated and asymmetric features have been detected toward both Class 0 \citep[e.g.,][]{pineda2020, cabedo2021, murillo2022, thieme2022}, Class I sources \citep[e.g.,][]{bianchi2022_siblings,valdivia-mena_2022, hsieh_2023, harada_2023, Podio_2024, Tanious_2024} and even more evolved system showing late infall \citep[e.g.,][]{gupta2024}. Molecular tracers not only enable the detection of these structures, but also provide insight into their physical and chemical properties. These observational results contrast with the traditionally symmetric picture of the protostellar phase \citep{shu1977, andre_2000, frank_2014}, indicating that protostellar accretion is often highly non axisymmetric and strongly influenced by the surrounding environment.

 {These structures, known as streamers, trace material funnelling from the outer envelope to the disk. They can be multiple and associated to disk misalignments, as the result of environmental interactions or captured material \citep[e.g.,][]{Gupta_2025}. Observed in molecular species as CO, HCO$^+$, HC$_3$N, they may contribute to protostellar accretion, disk substructure formation, and can carry enough mass to significantly replenish the disk \citep{pineda_ppvii_2023, Kuffmeier2023, hennebelle_2017,Kuznetsova2022,Tanious_2025} or alter its chemical composition through accretion shocks \citep{sakai_2014, Garufi2022,de_la_villarmois_2022, Podio_2024}.}

\begin{figure*} [htp!]
    \centering
    \includegraphics[width=0.85\textwidth]{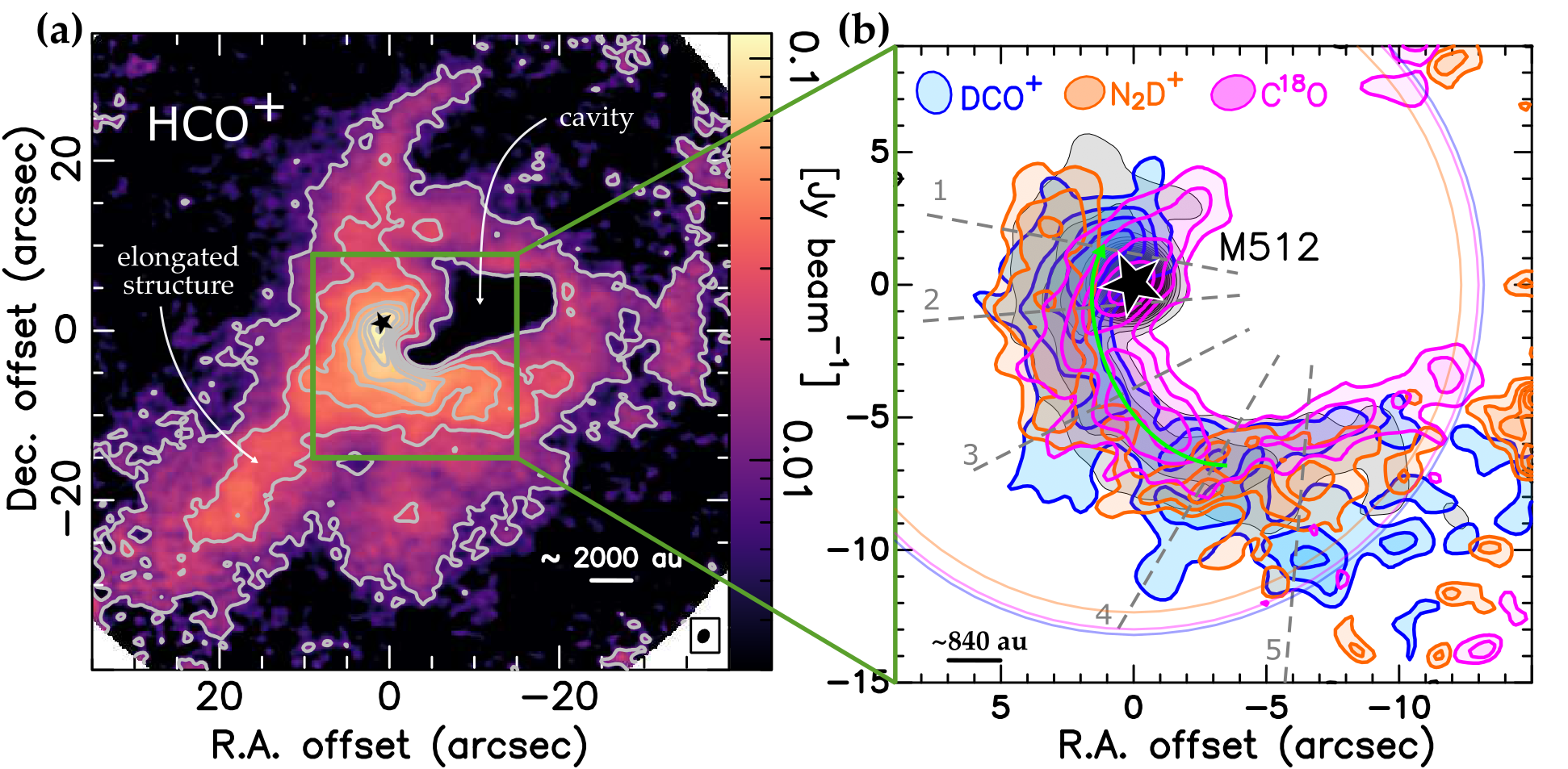}
    \caption{ {(a)}: HCO$^+$ emission toward M512.  {Beam is in the lower-right corner } {(b)}: Zoom-in on DCO$^+$ (blue), N$_2$D$^+$ (orange) and C$^{18}$O (magenta) moment 0, with 5$\sigma$ dust continuum (gray) from \citet{cacciapuoti24}. Separated maps are in Figure \ref{fig:app_mom0}.
    Dashed grey lines show the spatial profiles paths (Figure  \ref{fig:spatial_profile}).  {Beams are the colored filled circles, and the solid lines the primary beams. The green arrow is the TIPSY stream-model trajectory from \citet{cacciapuoti24}}.
     Contours starts at 3$\sigma$ with 2$\sigma$ steps ($\sigma$= 0.32, 0.24, 0.22 mJy beam$^{-1}$ km s$^{-1}$ for DCO$^+$, N$_2$D$^+$ and C$^{18}$O, respectively) and 5$\sigma$ steps for HCO$^+$ ($\sigma=$ 3 mJy beam$^{-1}$ km s$^{-1}$).
    }
    \label{fig:map_streamer}
\end{figure*}

One very peculiar elongated structures is toward the  {the low-mass Class I protostar M512}  in Orion A/Lynds 1641 \citep[$\sim$420 pc;][]{gaia_coll_2022}.  {This structure, detected in both gas and millimeter dust \citep{grant_2021}, was interpreted as a possible streamer by \citet{cacciapuoti24}, who used a streamline model of the C$^{18}$O emission and found that it is tentatively consistent with infall, deriving a mass of $\sim$80 M$_\mathrm{J}$ and a mass infall rate of 2 $\times$ 10$^{-6}$ M$_\odot$ yr$^{-1}$. However, their narrow ($\sim$1 km s$^{-1}$) line is only resolved by  4--5 channels in that dataset, preventing a robust kinematic characterization.    If even a fraction of this material reaches the disk, it could significantly affect its physical and chemical evolution, making it important to understand the structure's nature. Further details on the source are in Appendix \ref{app:source_descrip}.}


In this work, we investigate the emission of { HCO$^+$}, DCO$^+$, N$_2$D$^+$, and C$^{18}$O to explore the morphology and chemistry of the M512 elongated structure.
The species distribution reveals chemical stratification and an offset between gas and dust peaks associated with the gas temperature and density structure.
This raises the question of whether  {this structure is an accretion flow channelling material onto the disk, or whether it mainly traces cloud material shaped by environmental effects.}
Understanding this will provide new insights into the interplay between cloud, disk, and environment.



\section{Observations}
M512 (RA: 05h40$'$13.789$''$, DEC: $-07$d32m16.02s) was observed with ALMA in Band 6 on December 2019 (2019.1.00951.S, PI: Grant, S.) and in Band 3 on January 2023 (program 2022.1.00126.S, PI: Wendeborn, J.). Both programs achieved a resolution of $\sim1\farcs2$, and maximum recoverable scales of $11\farcs7$ for Band 6 and $14\farcs1$ for Band 3. 
Further details on data calibration are in \citet{grant_2021} and \citet{cacciapuoti24}.
For this work we used the 12m data of 
C$^{18}$O, HCO$^+$, DCO$^+$, and  N$_2$D$^+$  {(see Table \ref{tab:spec_par_lines})}. 
The two latter belongs to a low spectral resolution spectral window (channel width $\sim$5 km s$^{-1}$; see  {Table} \ref{tab:spec_par_lines}) and their detection is limited to one single channel. 
The cubes have been self-calibrated mapping the continuum-based gain solutions found in \citet{cacciapuoti24} to the line spectral windows closest in frequency, continuum subtracted and primary beam corrected. They were imaged in \texttt{CASA}\footnote{ {Common Astronomical Software Applications  { \citep{casa_team_2022}}:} https://casa.nrao.edu/} using \texttt{tclean} with a {natural} weighting, multiscale deconvolution (scales = {[0, 5, 15, 30])} {with a pixel size of $0.2''$}, and automasking. The resulting synthesized beams and rms are reported in  {Table} \ref{tab:spec_par_lines}. 
We analyzed the cube and produced moment maps using \texttt{CARTA}\footnote{{ {Cube} Analysis and Rendering Tool for Astronomy \citep{comrie_2026, wang_CARTA_2026}:} \url{https://carta.almascience.nrao.edu/}}, GILDAS\footnote{ {Grenoble Image and Line Data Analysis Software \citep{gildas_2013}:} \url{https://www.iram.fr/IRAMFR/GILDAS/}} and in-home python routines. We assume a 10\% flux calibration uncertainty \citep{Remijan2019}.

\section{Analysis and Results} \label{sec:Analysis-Results}

\begin{figure*} [htp!]
    \centering
\includegraphics[width=0.8\textwidth]{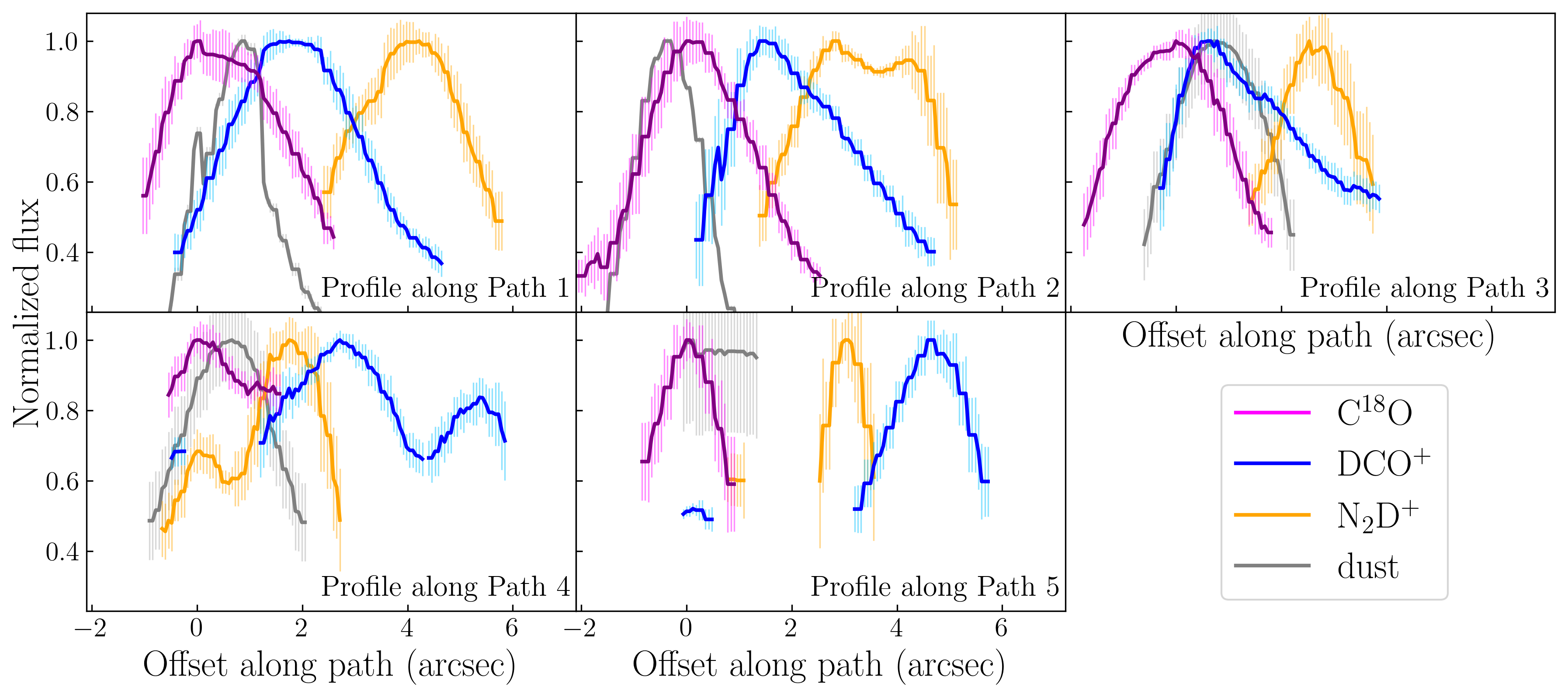}
    \caption{Normalized continuum (gray), C$^{18}$O (magenta), DCO$^+$ (blue), and N$_2$D$^+$ (orange) profiles along 5 cuts across the streamer ( {Figure} \ref{fig:map_streamer}). Fluxes are averaged over 3 pixels; error bars are  3$\times$3-pixel standard deviation. Offsets are relative to the C$^{18}$O peak.
    }
    \label{fig:spatial_profile}
\end{figure*}

\begin{table*}[]
    \centering
    \resizebox{\linewidth}{!}{%
    \begin{threeparttable}
    \caption{Spectroscopic and image parameters of the detected lines}
    \label{tab:spec_par_lines}
    \begin{tabular}{c|ccccccccc}
    \hline
    \hline
    Line & Transition & Frequency$^a$ & $\log_{10}$(A$_{ij}$)$^a$ & E$_{up}^a$ & g$_{up}^a$ & Spect. Res. & Beam & RMS & ALMA project \\
    &  & [GHz] &  & [K] & & km s$^{-1}$ & $[''\times '' (^\circ)]$ & [mJy/beam ] \\
    \hline
        C$^{18}$O & 2--1 & 219.56036 & -6.2 & 16 & 5 & 0.16 & 1.67$\times$ 1.26 (-72) & 13.2 & 2019.1.00951 \\
        N$_2$D$^+$ & 3--2 & 231.32183 & -3.8 & 22 & 63 & 5.1 & $1.5\times1.2$ (-74) & 2.8 &2019.1.00951  \\
        DCO$^+$ & 3--2 & 216.11258 & -3.1 & 21 & 7 & 5.4 & $1.28\times1.61$ (-75) & 3.5 & 2019.1.00951\\
        HCO$^+$ & 1--0 & 89.18852   &  -4.4 & 5 & 3 & 0.24 & $1.41\times1.75$ (62) & 1.7 &2022.1.00126 \\
    \hline
    \end{tabular}
     \begin{tablenotes}
         \footnotesize \item[$^a$] Spectroscopic parameters are by \citet{tinti_2007} for HCO$^+$, \citet{Winnewisser_1985} for C$^{18}$O, \citet{amano_2005, pagani_2009} for N$_2$D$^+$, \citep{caselli_2005, lattanzi_2007}  for DCO$^+$, from the CDMS catalog \citep{muller_cdms_2005}.
     \end{tablenotes}
       \end{threeparttable}
        }
\end{table*}


 {Figure}  \ref{fig:map_streamer}(a) shows the large scale moment 0 emission (integrated intensity)  {of HCO$^+$ over 2.4--4 km s$^{-1}$}. 
As expected for relatively abundant species  {and their lowest transition}, the emission seems to be dominated by the envelope. In addition to the central curved feature associated with the streamer identified by \citet{cacciapuoti24}, HCO$^+$ also traces an elongated structure to the south east and a cavity in the opposite direction. The envelope dominated nature of HCO$^+$ is further supported by the comparison between HCO$^+$ and C$^{18}$O channel maps shown in Appendix \ref{app:co_vs_hco+}. The streamer feature identified by \citet{cacciapuoti24} is clearly traced by C$^{18}$O over a narrow velocity range ($<1$km s$^{-1}$), spread across a few channels, indicating it traces the densest part of the structure. In contrast, HCO$^+$ is more extended and peaks close to the systemic velocity ($\sim$ 3.3--3.5 km s$^{-1}$), where strong envelope emission is expected. 
Therefore, the streamer signature in HCO$^+$ is clearly contaminated by the envelope, while C$^{18}$O provides a cleaner view of its kinematics.


In the inner 25$"$,  {Figure}  \ref{fig:map_streamer}(b) shows the moment 0 maps of C$^{18}$O,  {integrated over 2.8--4 km s$^{-1}$,} and  DCO$^+$, and N$_2$D$^+$, { integrated over the single channel of emission},  overlaid on the dust continuum. The same maps can be seen separately in  {Figure}  \ref{fig:app_mom0}. All three species show a similar arc-like structure following the dusty streamer from \citet{cacciapuoti24}  {and they show a clear stratification}: C$^{18}$O is primarily on the western side { on the eastern side of the cavity}, N$_2$D$^+$ is farther and toward the east, and DCO$^+$ seems to lie in between for most part of the streamer. The emission extends over $\sim$25$"$ ($\sim$10500 au) along the arc, with widths of $\sim$3--5$"$ (1200--2000 au depending on the tracer), reaching a maximum projected distance of $\sim$17$"$ ($\sim$7000 au) from the protostar.
To better characterize the stratification, we extracted spatial profiles along cuts intersecting the molecular emission peaks and oriented to highlight the differences between tracers ( {Figure}  \ref{fig:spatial_profile}). 
The cuts ( {Figure}  \ref{fig:map_streamer}b) have been chosen to follow the streamer's dust curvature rather than being perpendicular to the streamer or radial to the protostar, to better trace its morphology.
Flux at each position along the cuts was computed as the mean intensity along three pixels ($0\farcs6$, $\sim$250 au) perpendicular to the path, while the associated errors correspond to the standard deviation within a 3$\times$3 pixel box around each point. The profiles are normalized by their maximum value to emphasize the relative spatial variation, enabling direct comparison of the tracers.
The profiles extracted along these cuts ( {Figure}  \ref{fig:spatial_profile}) confirm the chemical stratification: in the three north-east cuts ( {1, 2, 3}), DCO$^+$ lies between C$^{18}$O and N$_2$D$^+$, whereas in the  {two} south-west cuts ( {4, 5}), N$_2$D$^+$ is located between C$^{18}$O and DCO$^+$.  {Note that, DCO$^+$ emission in the cut 4 and 5 lie near the edge of the primary beam, where sensitivity decreases, and looks fragmented} so the observed profiles there should be interpreted with caution. In addition, the apparent inversion of DCO$^+$ and N$_2$D$^+$  {could also be a result of projection effects not knowing the 3D structure}. 
Nevertheless, the overall stratification is evident, indicating variations in chemical composition and/or physical conditions across the streamer.



Given the detection of both HCO$^+$ and DCO$^+$ along the streamer, we estimated the deuteration ratio ( {N(DCO$^+$)/N(HCO$^+$)}) to probe chemical variations along the structure. This estimate, however, is subject to significant caveats: HCO$^+$ is likely optically thick and contaminated by the envelope, and DCO$^+$ is spectrally diluted due to the low spectral resolution preventing a determination as to whether both species trace the same gas.
To compute the ratio, we derived the column densities of both species, under the assumptions of Local Thermal Equilibrium (LTE) and optically thin emission (eq. \ref{eq.colDens}). We assumed a 10--30 K temperature range, {for a cold dense cloud ($\sim$10 K) and warmer gas in streamer (up to 30 K; \citealt{codella2024_streamer})}, in order to account the spectroscopy of two transitions ( {Table} \ref{tab:spec_par_lines}).
To minimize envelope contamination, the HCO$^+$ column density was estimated integrating the emission overlapping with C$^{18}$O (2.9–3.5 km s$^{-1}$;  {Figure}  \ref{fig:chmap_hco_vs_co}). 
 {We derived N(HCO$^+$) and N(DCO$^+$) along the streamer of 0.2--2 $\times$10$^{13}$ cm$^{-2}$ and 0.3--0.9 $\times$10$^{13}$ cm$^{-2}$, respectively. }
 { {HCO$^+$} becomes optically thick in the 10--30 K range with column density larger than 1$\times$10$^{13}$, so the derived column densities should be treated as lower limits, and the resulting ratio as an upper limit. }
We estimated N(DCO$^+$)/N(HCO$^+$) of $\sim$0.2--5, depending on temperature (10--30 K) and position along the streamer (Figure \ref{fig:hco+_dco+_ratio} at 30 K). These values should be taken with caution as they are  are upper limits for the reasons mentioned above. Additionally, deuteration seems enhanced in the south, where DCO$^+$ shows stronger emission (see  {Figure}  \ref{fig:map_streamer}, \ref{fig:hco+_dco+_ratio}). However this happens at the edge of the primary beam.
For completeness we computed the column density of N$_2$H$^+$ and C$^{18}$O (see Appendix \ref{app-sec:coldens}). 
 {Although HCO$^+$/DCO$^+$ could be used to estimate the ionization fraction and the cosmic ray ionization rate \citep[e.g.,][]{pineda_2024}, the \citet{caselli_1999} method requires cold $\sim$10 K pre-stellar conditions and co-spatial tracers, which are not met here, so the derived values would be overestimated \citep{sabatini_2023}.} 

\section{Environmental Shaping of the M512 Streamer}
\begin{figure}
    \centering
    \includegraphics[width=0.75\linewidth]{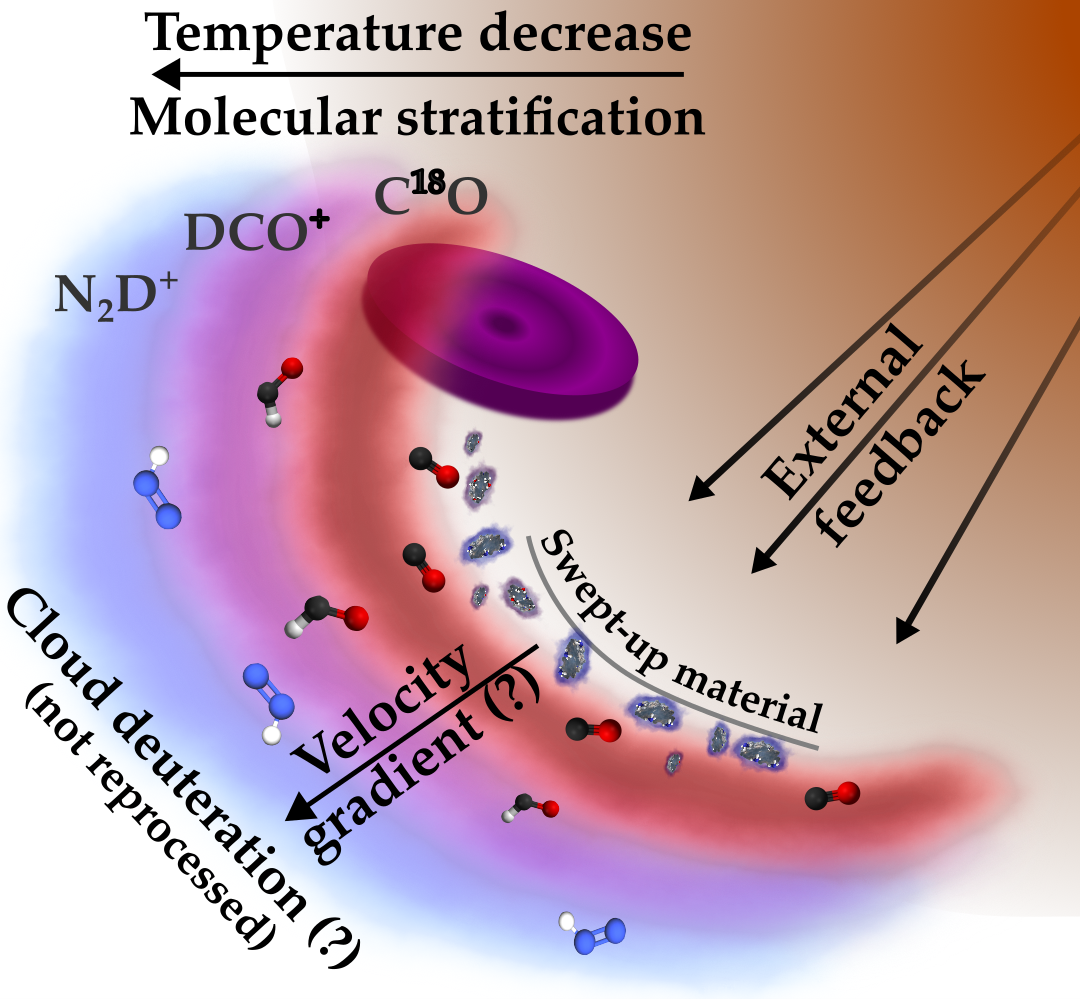}
    \caption{
    Scheme of the proposed external feedback  scenario shaping the M512 structure. N$_2$D$^+$ (blue), traces the cold outer regions, DCO$^+$ (purple) the intermediate layer, and C$^{18}$O (red) the warmer gas. The M512 protostar/disk is at the center. Arrows indicate feedback direction and the resulting velocity and temperature gradients. The figure is qualitative and not to scale. } \label{fig:scheme}
    \label{fig:placeholder}
\end{figure}

The molecular stratification observed along the streamer provides insights into its physical and chemical structure. The arrangement of N$_2$D$^+$ in the outer eastern regions, DCO$^+$ in the middle, and C$^{18}$O on the west close to the cavity 
reflects the interplay between temperature, CO depletion, and deuterium chemistry. In particular, the enhanced outer N$_2$D$^+$ indicates CO depletion (that happens below 25 K), DCO$^+$ forms where some CO remains and where N$_2$D$^+$ and CO coexists \citep{caselli_deuterated_2002,pagani_2011,ceccarelli2014ppvi}, and the inner C$^{18}$O reflects regions where CO has sublimated. Dust accumulation near the eastern edge may enhance CO shielding and the increase of local density. 
 {The presence of C$^{18}$O emission  at the eastern edge of the streamer cannot be explained by t {he protostar}: with $\sim 1$ L$_\odot$, at $\sim$7000 au the dust temperature is below the CO sublimation temperature (20–25 K), and CO should be frozen onto dust grain. Detecting gas-phase CO at this location suggests the presence of an additional local heating source, such as compression or external irradiation.}

 {The chemical stratification and the dust morphology identified in our analysis suggest that the elongated structure could have been shaped by external feedback, potentially pushing material from north-east (NE) to south-west  (SE). These signatures naturally arise in a swept‑up layer that produces dust accumulation, a localized temperature increase which release C$^{18}$O, and the preservation of deuterated species in the outer regions, and they are consistent with the tentative NE–SW velocity gradient seen in C$^{18}$O \citep{cacciapuoti24}.
In this scenario, shown in Figure \ref{fig:scheme}, the observed molecular layering and morphology are more naturally explained than in the alternative interpretations discussed by \citet{grant_2021} and \citet{cacciapuoti24} (see Section~\ref{app:source_descrip}).
Also, the lack of FIR nebulosity and the normal Class‑I–like SED (see section \ref{app:source_descrip}) suggest that any external influence is mild, consistent with the moderate heating or compression required to produce the observed stratification.  
Filamentary and streamer-like structures are ubiquitous in star-forming regions, but their formation and role is still discussed \citep[e.g.,][]{hennebelle_2019, robitaille_2020}. Often they are influenced by external triggers such as expanding bubbles, shocks, or nearby stellar activity \citep[e.g.,][]{hacar_orion_2018, socci_fiberOrion_2024, De_simone_train_2022, chahine_omc-2_2022}. 
Orion shows a filamentary environment with cavities and feedback‑driven structures \citep{Feddersen_COshells_2018, Zheng_symmetry_2021}, and intermediate FUV radiation fields from nearby A–B stars are known to influence the surrounding gas and disks \citep{vanTerwisga_Hacar_2023} even though the local FUV field around M512 is modest (Appendix \ref{app:source_descrip}). Note that the field covered by \citet{Feddersen_COshells_2018} do not include M512, but they show that small‑scale feedback is widespread in Orion A (Figure \ref{fig:orion_M512}).
In this view, the morphology and chemical stratification of the M512 structure are more naturally explained by environmental shaping than by a single coherent gravitational accretion flow. Only its densest inner region is likely to accrete onto the disk, while the bulk of the elongated feature reflects the imprint of the surrounding cloud, possibly channelled through the parent filament \citep{valdivia-mena_propstar_2024}, with limited physical or chemical impact on the disk itself.}


Higher‑resolution ($<$0.1 km s$^{-1}$) data are needed for kinematics, and slightly larger‑scale shock and dense‑gas mapping could probe gentle compression or illumination in the surrounding filament.


\begin{acknowledgements}
We thank the referee for helping improve the manuscript. We thank Prof. C. Ceccarelli and Dr. G. Sabatini for the useful discussions.
This paper makes use of the following ALMA data: ADS/JAO.ALMA\#2019.1.00951.S and ADS/JAO.ALMA\#2022.1.00126. ALMA is a partnership of ESO (representing its member states), NSF (USA) and NINS (Japan), together with NRC (Canada), NSTC and ASIAA (Taiwan), and KASI (Republic of Korea), in cooperation with the Republic of Chile. The Joint ALMA Observatory is operated by ESO, AUI/NRAO and NAOJ. D.C. acknowledge support from the 2024 ESO Summer Research Programme. 
\end{acknowledgements}

%
%
\bibliographystyle{aa}
\bibliography{stremM512} 

\begin{appendix}
\onecolumn

\section{Source description} \label{app:source_descrip}
 {M512 (also known as [MGM2012] 512 or 2MASS J05401378-0732160) is a young stellar object located in the Lynds 1641 region of the Orion A molecular cloud at a distance of $\sim$420 pc, based on the Gaia DR3 parallax (p $\sim$ 2$\farcs$38; \citealt{gaia_coll_2022}). Although the source has been identified as young star with a disk \citep{Megeath_2012}, \citet{Caratti_garatti_2012}  classify it as a Class I based on the 2--25 $\mu$m spectral index ($\alpha_{2-25 \mu m}\sim$ 0.33; ), with a stellar mass of 0.15 M$_\odot$ and a luminosity of 0.9 L$_\odot$ derived from SED modeling.
Extended millimeter continuum and CO isotopologue emission surrounding the disk were first reported by \citet{grant_2021}, revealing a large-scale elongated structure connected to the central source. The CO emission does not show clear signatures of outflow emission (through morphology and/or velocty structure) and it is primarily associated with the elongated structure, while the disk kinematics and orientation remain uncertain. The dust mass of the extended emission is estimated to be 65$\pm$32 M$_\oplus$, corresponding to about 50\% of the disk mass; if this material belongs to an envelope, the envelope-to-disk mass ratio would be consistent with a Class I source \citep{jorgersen_2009}, although the disk mass may be underestimated due to optical depth effects.
The origin of this elongated structure is uncertain. \citet{grant_2021} discussed several possible scenarios, including an envelope or accretion streamer, an outflow cavity wall, perturbations from a companion, or interaction with  the parent cloud. They concluded that the detection of the structure in dust continuum implies the presence of relatively large grains and dense material, and that interaction between the young system and the surrounding cloud may play an important role.
More recently, \citet{cacciapuoti24} modeled the structure as a streamer using C$^{18}$O emission and derived a mass of $\sim$80 M$_\mathrm{J}$ and a high infall rate of 2 $\times$ 10$^{-6}$ M$_\odot$ yr$^{-1}$, suggesting that, if fully accreted, this structure could significantly affect the disk and the protostar. However, the nature of the structure and the extent to which this material is effectively accreting onto the disk remain uncertain. In this context, studying the chemical structure of the elongated feature can provide important constraints on its physical origin and its role in the star-disk system.
To place these properties in context, it is important to consider the larger scale environment in which M512 is embedded. M512 lies within the southern part of Orion A, an environment that contains numerous cavities and feedback-driven structures on (sub-)parsec scales. Although the field mapped by \citet{Feddersen_COshells_2018} do not intersect this specific region, such structures illustrate that small-scale feedback is common across Orion A and may influence the local gas distribution around M512. Additionally, several A-B stars in L1641 contribute to a moderate FUV radiation field \citep{vanTerwisga_Hacar_2023}, but no individual irradiating source can be uniquely associated with M512. The local radiation field seems to be relatively weak  ($<$10 G$_0$) \citep{vanTerwisga_Hacar_2023}. 
Additionally, Herschel 70 $\mu$m maps show no extended nebulosity around M512 \citep{grant_2018}, indicating the absence of strong local heating or illuminated cavity walls on $\sim$10$^{3-4}$ au scales. Furthermore, the source's SED closely resembles that of a normal Class I protostar viewed at low inclination (\citealt{grant_2018}, with no excess emission that would suggest strong external irradiation. These FIR and SED characteristics imply that any environmental influence on the extended material must be relatively mild, acting primarily on the surrounding cloud rather than on the protostar itself.
While this rules out strong photo-evaporation or intense shock heating, it remains consistent with a scenario in which moderate external compression, low-level FUV illumination, or larger-scale cloud dynamics shape the morphology and chemistry of the material surrounding M512.}

\begin{figure}[hb!]
\centering
\includegraphics[width=0.95\textwidth]{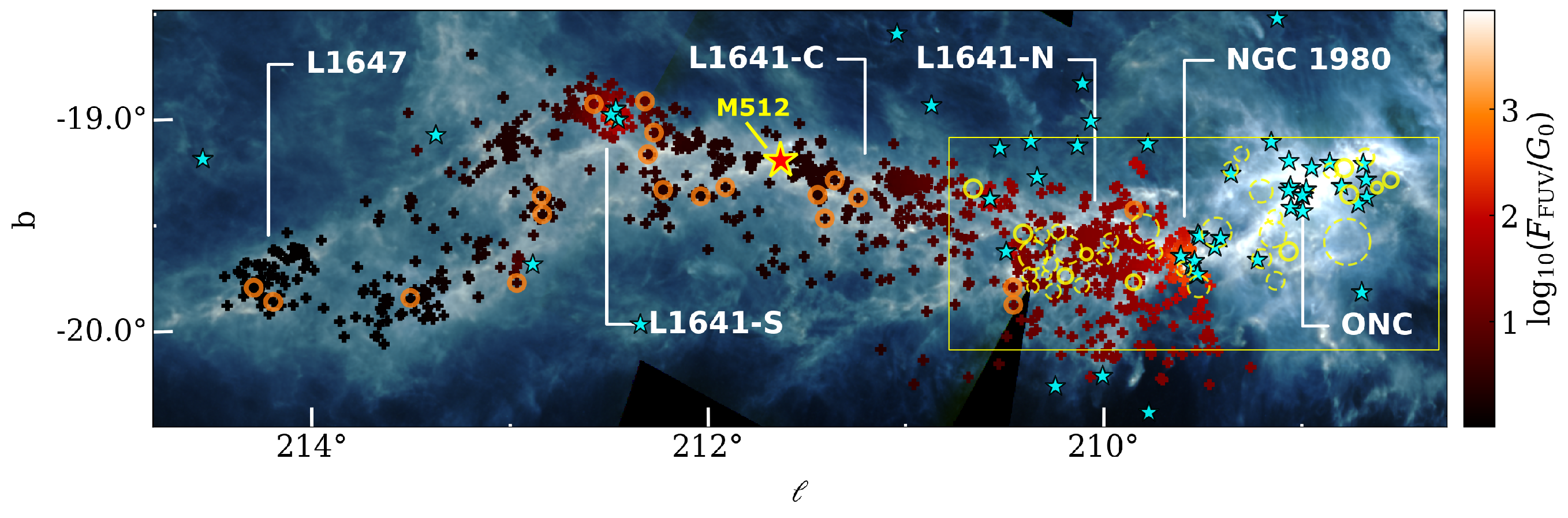}
\caption{ {Herschel SPIRE observations at 250, 300, and 500 $\mu$m (background) overlapped with external FUV flux estimated map in the Orion A cloud from \citet{vanTerwisga_Hacar_2023}. The cyan stars mark possible ionizing stars in their vicinity and crosses show the Class II disks. Active star-forming regions and the location of the M512 protostar are indicated. The yellow circles show the CO shells identified by \citet{Feddersen_COshells_2018} in the yellow square field.} }\label{fig:orion_M512}
\end{figure}

\newpage

\section{Channel maps of C$^{18}$O and HCO$^+$} \label{app:co_vs_hco+}
\begin{figure*}[b!]
    \centering
    \includegraphics[width=0.85\textwidth]{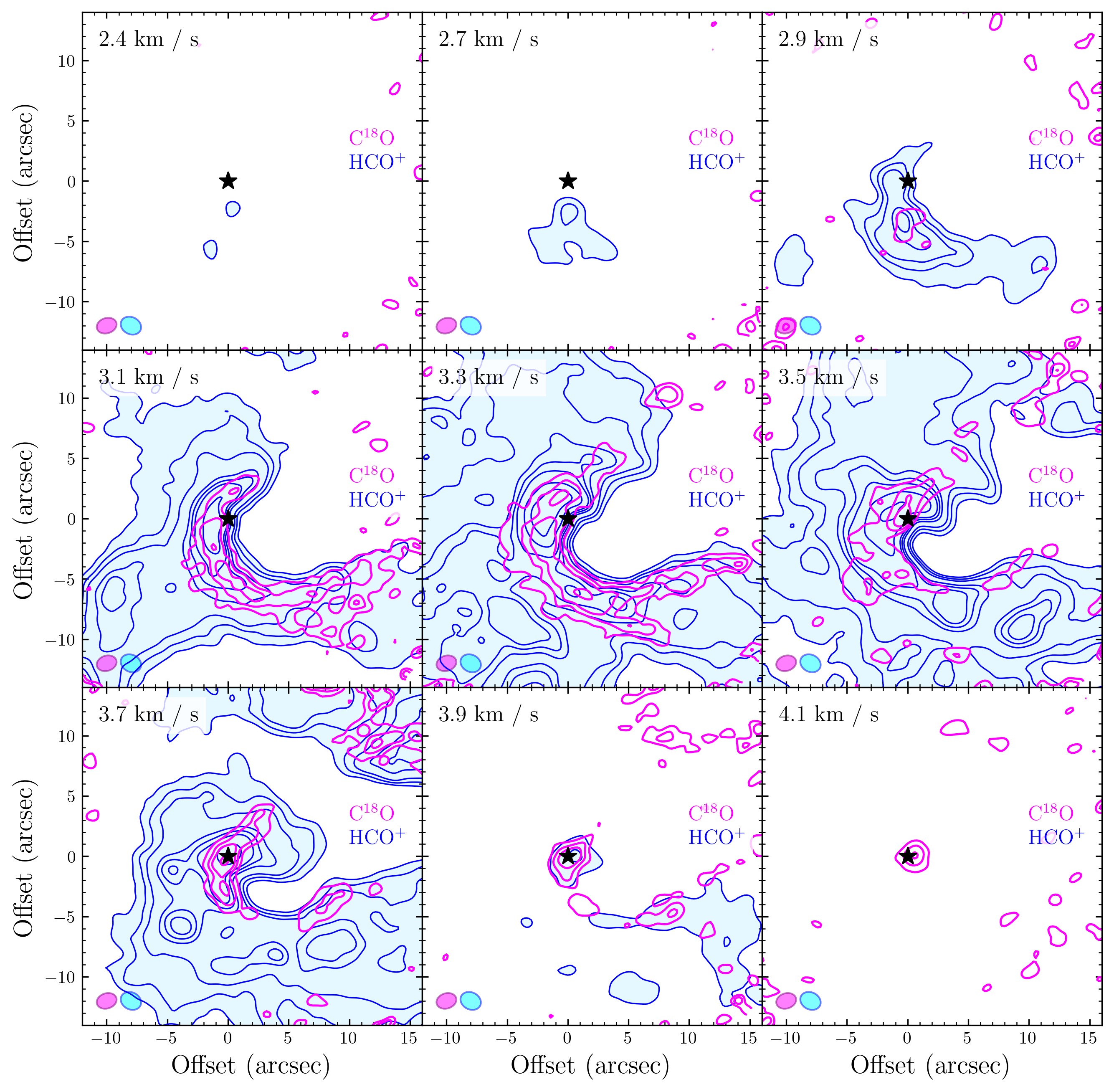}
    \caption{Kinematics of C$^{18}$O (magenta) and HCO$^+$ (blue) shown through channel maps. The synthesized beams are shown in the lower left corner. The velocity bins are depicted in the upper right corner. 
    Contours are [3, 5, 7, 11, 15] $\sigma$ for C$^{18}$O with $\sigma$=0.26 mJy beam$^{-1}$ and [3, 5, 7, 11, 13, 15, 20, 25, 30] $\sigma$ for HCO$^+$ with $\sigma=$5 mJy beam$^{-1}$. }
    \label{fig:chmap_hco_vs_co}
\end{figure*}

The channel maps in  {Figure}  \ref{fig:chmap_hco_vs_co} illustrate the kinematics of C$^{18}$O and HCO$^+$ around the M512 streamer. C$^{18}$O emission is confined to a narrow velocity range ($\sim$2.9–3.5 km s$^{-1}$), tracing the densest parts of the structure, while HCO$^+$ is more extended and peaks near the systemic velocity, indicating significant contribution from the surrounding envelope. Only a few channels show HCO$^+$ emission coincident with the streamer, highlighting the difficulty of isolating the streamer in this tracer due to spectral dilution. These maps therefore demonstrate that C$^{18}$O provides a cleaner probe of the streamer kinematics, while HCO$^+$ is better suited for tracing the overall envelope, supporting its use as a complementary rather than primary tracer.

\newpage
\section{Moment comparison of N$_2$D$^+$, DCO$^+$ and C$^{18}$O with dust emission} \label{app-sec:moments}
\begin{figure}
    \centering
    \includegraphics[width=0.9\linewidth]{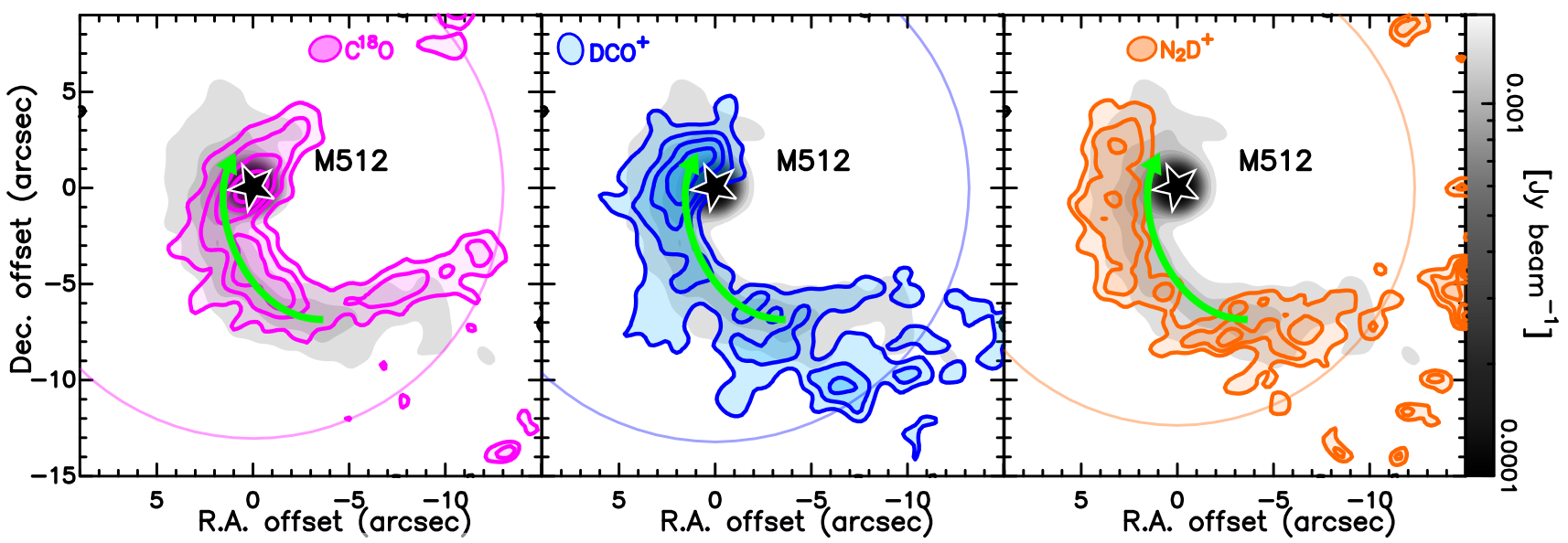}
    \caption{Same as  {Figure}  \ref{fig:map_streamer}(b) but shown in separated panels for clarity.  {The green arrow represent the TIPSY stream-model trajectory for 0.15 M$_\odot$ and v$_{sys}$ of 3.25 km s$^{-1}$ from \citet{cacciapuoti24}.}}
    \label{fig:app_mom0}
\end{figure}

 {Figure}  \ref{fig:app_mom0} shows the individual moment 0 maps of C$^{18}$O, DCO$^+$, and N$_2$D$^+$ over the same field as  {Figure}  \ref{fig:map_streamer}(b). Displaying the tracers separately highlights their different spatial extents and relative offsets with respect to the dust continuum, which are discussed in detail in Sect. \ref{sec:Analysis-Results}. These maps provide a complementary view to the overlaid representation shown in the main text.

\section{Column densities and abundance ratio}\label{app-sec:coldens}

The column densities were computed under the assumptions of Local Thermal equilibrium (LTE) and optically thin emission, using the integrated intensity ($\rm \int T_b Dv$) of each line as \citep{mangum_2015, goldsmith1999ApJ}:
\begin{equation}
\rm N_{\rm tot} =\frac{8\pi k \nu^2}{h c^3 A_{ul}}\frac{Q(T_{\rm rot})}{g_u} \exp\!\left(\frac{E_u}{kT_{\rm rot}}\right)\int T_{\rm b}\, dv
\end{equation} \label{eq.colDens}
where N$_{\rm tot}$ is the total column density, $k, \, h$ the Boltzmann and Plank constants, $c$ the speed of light, $\nu$ is the line rest frequency, $A_{ul},\, g_u,\, E_u$ the line spectroscopic parameters, $Q(T_{\rm rot})$ the partition function at rotational temperature $T_{\rm rot}$.

\begin{figure}[b]
    \centering
    \includegraphics[width=0.9\linewidth]{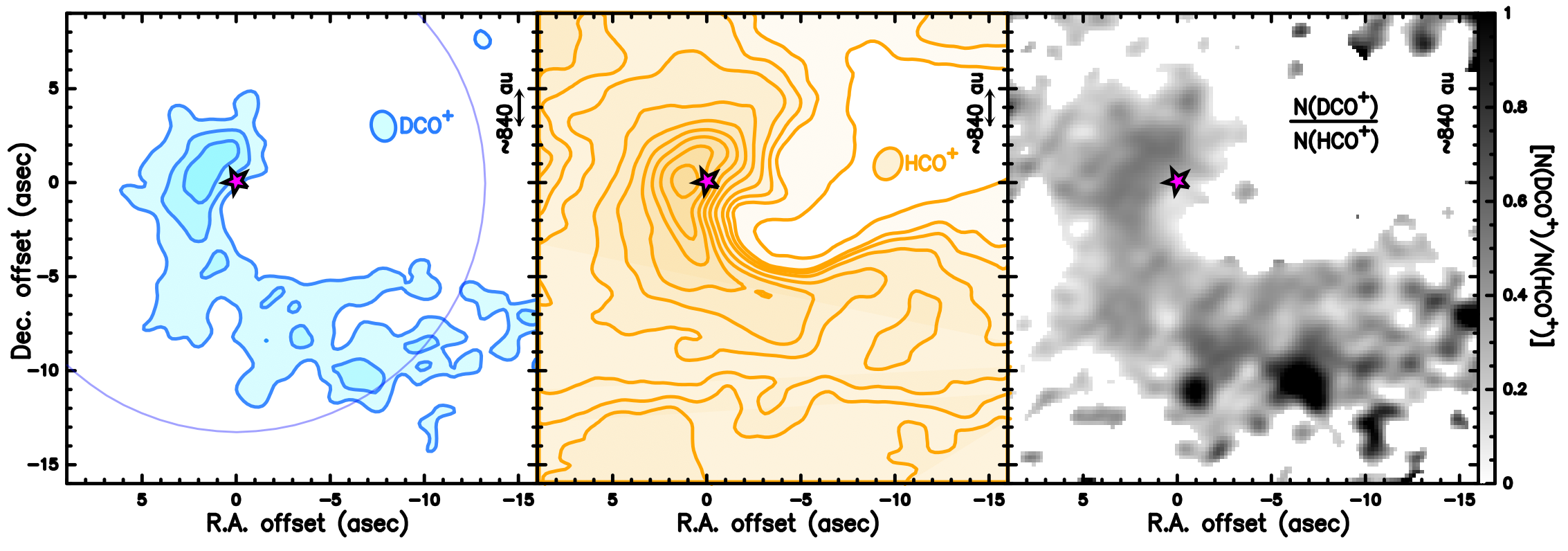}
    \caption{ {N(DCO$^+$)/N(HCO$^+$)} column density ratio computed assuming 30 K (right panel) and using the moment 0 intensity map of HCO$^+$ (middle panel in orange) and DCO$^+$ (left panel in blue). Contours starts at 3$\sigma$ with 3$\sigma$ steps ($\sigma$=0.32 mJy beam$^{-1}$ km s$^{-1}$) for DCO+ and 6$\sigma$ steps ($\sigma$=3 mJy beam$^{-1}$ km s$^{-1}$) for HCO$^+$. Note that the ratio values represent an upper limit due to optical thickness and contamination of HCO$^+$ (see text). Primary beam and beams are reported.}
    \label{fig:hco+_dco+_ratio}
\end{figure}

 {Figure}  \ref{fig:hco+_dco+_ratio} shows the map of the column density ratio of DCO$^+$ and HCO$^+$ (values in Section \ref{sec:Analysis-Results}).  
We computed as well the column density of N$_2$D$^+$ and C$^{18}$O using the same assumption of LTE and optically thin emission, in the 10--30 K temperature range. 
We found N(N$_2$D$^+$) of 3--8 10$^{11}$ cm$^{-2}$
and N(C$^{18}$O) of 0.7--1.7 10$^{14}$ cm$^{-2}$ along the streamer. 

\end{appendix}

\end{document}